\documentclass[letterpaper,twocolumn,prl,aps,superscriptaddress,floatfix]{revtex4-2}

\usepackage[T1]{fontenc}
\usepackage[latin9]{inputenc}
\usepackage{amsmath}
\usepackage{amssymb}
\usepackage{stackrel}
\usepackage{setspace}
\usepackage{xcolor}

\makeatletter

\usepackage{siunitx}
\DeclareSIUnit{\belmilliwatt}{Bm}
\DeclareSIUnit{\dBm}{\deci\belmilliwatt}
\DeclareSIUnit{\Henry}{H}
\DeclareSIUnit{\pH}{\pico\Henry}
\DeclareSIUnit{\Kelvin}{K}
\DeclareSIUnit{\mK}{\milli\Kelvin}

\usepackage[unicode=true,pdfusetitle,
 bookmarks=true,bookmarksnumbered=false,bookmarksopen=false,
 citecolor=cyan,urlcolor=magenta,linkcolor=blue, breaklinks=false,
 pdfborder={0 0 0},backref=false,colorlinks=true]{hyperref}

\usepackage{amsthm}\usepackage{epstopdf}\usepackage{color}
\usepackage{bm}\usepackage{babel}
\usepackage{upgreek}
\usepackage{graphicx}
\usepackage{siunitx}
\makeatother

\usepackage{babel}
\usepackage{newtxtext,newtxmath}  

\begin{document}

\title{Hybridized Frequency Combs in Multimode Cavity Electromechanical System}

\author{Sishi Wu}
\affiliation{Quantum Physics and Quantum Information Division, Beijing Computational
Science Research Center, Beijing 100193, China}

\author{Yulong Liu}
\email{liuyl@baqis.ac.cn}
\affiliation{Beijing Academy of Quantum Information Sciences, Beijing 100193, China}

\author{Qichun Liu}
\affiliation{Beijing Academy of Quantum Information Sciences, Beijing 100193, China}

\author{Shuai-Peng Wang}
\affiliation{Quantum Physics and Quantum Information Division, Beijing Computational
Science Research Center, Beijing 100193, China}

\author{Zhen Chen}
\affiliation{Beijing Academy of Quantum Information Sciences, Beijing 100193, China}

\author{Tiefu Li}
\email{litf@tsinghua.edu.cn}
\affiliation{School of Integrated Circuits and Frontier Science Center for Quantum Information, Tsinghua University, Beijing 100084, China}
\affiliation{Beijing Academy of Quantum Information Sciences, Beijing 100193, China}
\date{\today}

\begin{abstract}
The cavity electromechanical devices with radiation-pressure-interaction induced Kerr-like nonlinearity are promising candidate to generate microwave frequency combs. We construct a silicon-nitride-membrane based superconducting cavity electromechanical device and study two mechanical modes mediated synergistic frequency combs. Around the threshold of intracavity field instability, we firstly show independent frequency combs with tooth spacing equalling to each mechanical mode frequency. At the overlap boundaries of these two individual mechanical mode mediated instability thresholds, we observe hybridization of frequency combs based on the cavity field mediated indirect coupling between these two mechanical modes. The spectrum lines turn to be unequally spaced, but can be recognized into combinations of the coexisting frequency combs. Beyond the boundary, the comb reverts to the single mode case, and which mechanical mode frequency will the tooth spacing be depends on the mode competition. Our work demonstrates mechanical mode competition enabled switchability of frequency comb tooth spacing and can be extended to other devices with multiple nonlinearities.
\end{abstract}

\maketitle

Frequency combs are spectra of phase-coherent evenly spaced narrow lines and initially invented in laser for frequency metrology~\cite{fortier201920,karim2020ultrahigh}. Originally, the optical frequency combs are developed in 1990's as rulers to measure large optical frequency differences and to provide a direct link between optical and microwave frequency~\cite{kourogi1993wide,diddams2000direct}. After two decades of development, apart from the application in metrology, the frequency combs have become powerful tool in molecular fingerprinting~\cite{diddams2007molecular}, atomic clocks~\cite{gohle2005frequency}, attosecond science~\cite{tzallas2003direct}, optical communications~\cite{torres2014optical,choudhary2017chip,gaeta2019photonic}, etc. Towards further application on quantum information and quantum computation, intrinsic phase matching of frequency comb makes it a promising platform for realizing multipartite entanglement and cluster states, which are building blocks in continuous variable quantum computation~\cite{roslund2014wavelength,chen2014experimental,tian2019quantum,pfister2019continuous,cai2020versatile}. 

In addition to atomic-gas based optical frequency combs, microcavities with Kerr nonlinearity~\cite{kippenberg2011microresonator,fujii2017effect,xu2019microcavity} have also attracted intensive attention. To explore the microwave frequency combs, electromechanical devices can be a suitable platform. They have relative strong single-photon couplings~\cite{teufel2011circuit,pirkkalainen2013hybrid,heikkila2014enhancing,pirkkalainen2015squeezing}, and can work at milli-Kelvin temperature, which makes them well isolated from most thermal noises environment. Meanwhile, sideband cooling technology~\cite{yong2013review,gan2019intracavity,lai2020nonreciprocal} enables electromechanical devices to be working in the quantum regime. Besides, capability of its integration with microwave superconducting circuits makes it a considerable composite for hybrid quantum computing system~\cite{clerk2020hybrid}. Here, we report on the observation of microwave frequency combs in a multimode electromechanical system composed of a three-dimensional (3D) aluminum superconducting cavity and a metalized silicon nitride (SiN) membrane.  

Besides each single-mechanical-mode mediated frequency combs with repetition rate of the respective characteristic frequency, we observe finer spectra, but with unequally spaced distribution. Such spectra result from hybridization of single mechanical mode mediated frequency combs, and occur only at overlap boundary of pump conditions for two individual mechanical induced cavity field instability thresholds. Thus, in our experiment, the pump frequency and power can be used to in-situ switch the frequency combs with different tooth spacing.

\begin{figure}[tb]
\includegraphics[width=8cm]{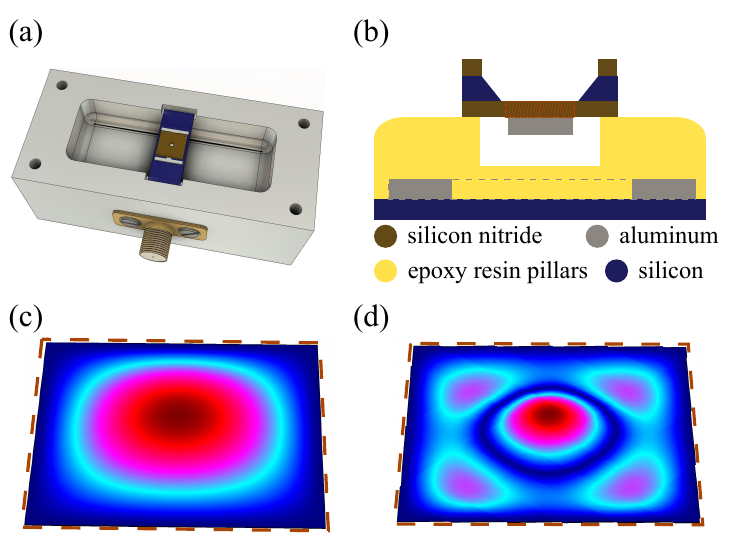}\\[5pt]
\caption{(a) Illustration of sample encapsulation. The mechanically compliant capacitor chip is placed at the center of the rectangle Al-box. (b) Side view cartoon of the mechanically compliant capacitor chip. The color indicates different materials. (c-d) Simulated motion of the first and the second mechanical modes (with a drumhead vibration at the membrane center) of the mechanical oscillating element, respectively.
}
\label{fig:1} 
\end{figure}

\textit{The Multimode Cavity Electromechanical Device.---}As is presented in Fig.~\ref{fig:1}(a), the sample is composed of a SiN-membrane based mechanical compliant capacitor and a 3D aluminium (Al) cavity. The SiN membrane holds multiple mechanical modes with different mode shape. Here, we consider the lowest two mechanical modes with membrane center drumhead vibration frequencies of $\Omega_{\mathrm{m1}}$ and $\Omega_{\mathrm{m2}}$, respectively. The internal chamber of the rectangle Al block supports the cavity mode, with a resonance frequency  of $\Omega_{\mathrm{c}}$. Figure~\ref{fig:1}(b) gives detailed information of the mechanically compliant capacitor chip. Two large Al coupling capacitors (gray parts on the bottom chip) laying on the high resistance silicon substrate (blue) ensure maximum single photon coupling strength while introducing the minimum stress disturbance for SiN membrane~\cite{liu2021gravitational,liu2021optomechanical}. High-stress SiN films (brown parts on the flip chip) are evaporated, with $20~\mathrm{nm}$ Al electrode on its back (gray). The reddish brown shaded area represents the suspend SiN membrane, supporting the mechanical oscillating part. The mechanical modes are simulated via finite element method, leading to expected eigen mode frequencies of $755~\mathrm{kHz}$ and $1.755~\mathrm{MHz}$, as is shown in Fig.~\ref{fig:1}(d) and (e), respectively.

Our experiment is carried out in a dilution refrigerator~\cite{SM}. The pump signal ($\Omega_{\mathrm{d}}$) is injected into the cavity through a circulator. Before reaching the sample, input signal is attenuated to reduce background noise. The reflected signal ($\Omega_{\mathrm{s}}$) is split into two, one of which is read by network analyzer, and the other is used to do the spectrum analyze. To avoid amplifier saturation, reflected pump signal is well cancelled by using a directional coupler~\cite{SM}.

Working in a frame rotating at the frequency of the pump tone, the Hamiltonian of such multi-mechanical-mode electromechanical system can be expressed as~\cite{aspelmeyer2014cavity,miri2018optomechanical,kemiktarak2014mode,buchmann2015nondegenerate,jiang2021energy,ockeloen2019sideband}, 
\begin{align}
H_{\mathrm{sys}}/\hbar=& (-\Delta_{\mathrm{dc}}-i\frac{\kappa}{2})a^{\dagger}a\nonumber + \stackrel[\mathrm{j}=1,2]{}{\Sigma} (\Omega_{\mathrm{mj}}-i\frac{\gamma_{\mathrm{j}}}{2})b_{\mathrm{j}}^{\dagger}b_{\mathrm{j}}\nonumber \\
 & + \stackrel[\mathrm{j}=1,2]{}{\Sigma} g_{\mathrm{j}}a^{\dagger}a(b_{\mathrm{j}}+b_{\mathrm{j}}^{\dagger}) + i\sqrt{\kappa_{\mathrm{e}}}S_{\mathrm{in}}(a^{\dagger}-a) ~,
\label{eq:1}
\end{align}

\noindent where $\Delta_{\mathrm{dc}}=\Omega_{\mathrm{d}}-\Omega_{\mathrm{c}}$, and foot label `$j$' in $\Omega_{\mathrm{mj}}$ indicates the order of the mechanical mode referred. Coefficient $g_{\mathrm{j}}$ represents the single photon coupling strength between cavity and the $j$-th mechanical mode, respectively. Coupling efficiency between the SMA connector and the cavity is labelled as $\kappa_{\mathrm{e}}$. $S_{\mathrm{in}}$ describes amplitude of the pump signal normalized to a photon flux at the input of the cavity, and are defined as $S_\mathrm{in}=\sqrt{P_\mathrm{d}/\hbar\Omega_\mathrm{d}}$, where $P_\mathrm{d}$ (in the unit of watt) is the pump power at the input of the cavity. Besides, $a$ ($a^{\dagger}$) and $b_{\mathrm{j}}$ ($b_{\mathrm{j}}^{\dagger}$) represent annihilation (creation) operators for cavity and the $j$-th mechanical mode, respectively. 

According to pre-characterization by measuring the mechanical spectra~\cite{SM}, frequencies of mechanical oscillator are $\Omega_{\mathrm{m1}}/2\pi=756~\mathrm{kHz}$ for the first mode (mode-1), with damping rate $\gamma_{\mathrm{1}}/2\pi=2.32~\mathrm{Hz}$ and $\Omega_{\mathrm{m2}}/2\pi=1.750~\mathrm{MHz}$ for the second mode (mode-2), with damping rate $\gamma_{\mathrm{2}}/2\pi=0.30~\mathrm{Hz}$. The 3D cavity is characterized to have frequency of $\Omega_{\mathrm{c}}/2\pi=5.31~\mathrm{GHz}$ and damping rate of $\kappa/2\pi=380~\mathrm{kHz}$, which implies that the electromechanical system is within the resolved-sideband regime ($\Omega_{\mathrm{mj}}>\kappa$)~\cite{aspelmeyer2014cavity,miri2018optomechanical,kemiktarak2014mode,buchmann2015nondegenerate,jiang2021energy,ockeloen2019sideband,schliesser2008resolved,liu2018cavity}. 
The single photon coupling strength $g_1/2\pi=0.49~\mathrm{Hz}$ and $g_2/2\pi=0.07~\mathrm{Hz}$ are calibrated from the measurements based on frequency modulation technique~\cite{SM}.

As implied by the system Hamiltonian, interaction of cavity and membrane results in photon number dependent cavity frequency shift, which in turn changes reflection. The equations of motion are

\begin{subequations}
\begin{eqnarray}
\dot{a}=\{i[\Delta_{\mathrm{dc}}-\stackrel[\mathrm{j}=1,2]{}{\Sigma} g_{\mathrm{j}}(b_{\mathrm{j}}+b_{\mathrm{j}}^{\dagger})]-\frac{\kappa}{2}\}a+\sqrt{\kappa_{\mathrm{e}}}S_{\mathrm{in}}~, \label{eq:2a}
\end{eqnarray}
\begin{eqnarray}
\dot{b_{\mathrm{j}}}=-(i\Omega_{\mathrm{mj}}+\frac{\gamma_{\mathrm{j}}}{2})b_{\mathrm{j}}-ig_{\mathrm{j}}a^{\dagger}a~.\label{eq:2b}
\end{eqnarray}
\end{subequations}

Similar to the frequency comb induced by Kerr oscillator~\cite{PhysRevA.80.065801,aldana2013equivalence,liu2017controllable}, such electromechanical couplings can also result in frequency comb for pump condition beyond threshold of intra-cavity field instability~\cite{miri2018optomechanical}, whereas the cavity field can be expressed as weighted sum of series Bessel functions~\cite{kemiktarak2014mode,SM}.

\begin{figure}[tb]
\includegraphics[width=8cm]{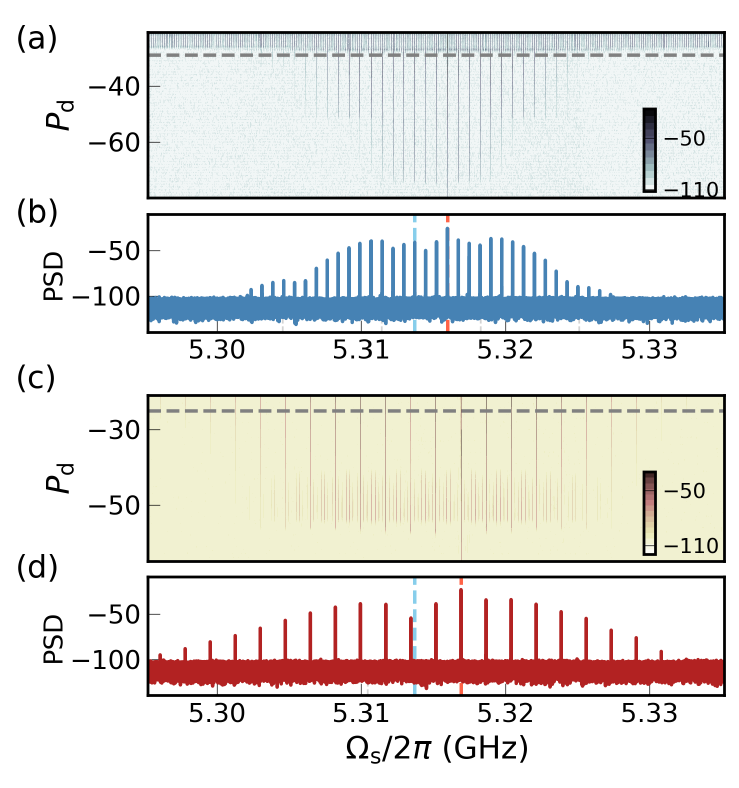}\\[5pt] 
\caption{Single-mechanical-mode mediated frequency combs. (a,c) Detected comb evolution along increased pump power with detuning of $\Delta_{\mathrm{dc}}\approx\Omega_{\mathrm{mj}}$. (b,d) An example of comb spectrum, and is marked as gray dash line in the respectively upper one. The pump power $P_\mathrm{d}$ is in the unit of $\mathrm{dBm}$. Unit of the power spectral density (PSD) is $\mathrm{dBm/Hz}$. The pump frequency is marked as vertical red dash line, and cavity frequency is marked as vertical blue dash line.}
\label{fig:2}
\end{figure}

\textit{Formation and Evolution of Electromechanical Frequency Combs.---}Firstly, we explore the formation of frequency combs under the exact blue sideband pump. The detuning between pump frequency to the cavity frequency matches one of the mechanical mode frequencies. Therefore, the cavity nonlinearity will be mainly determined accordingly by one mechanical motion. We then conduct the heterodyne detection~\cite{SM} of the reflected spectrum, whose x-axis is labelled as $\Omega_\mathrm{s}$. When $\Delta_{\mathrm{dc}}=\Omega_{\mathrm{m1}}$, as shown in Fig.~\ref{fig:2}(a), only one peak at $\Omega_{\mathrm{s}}=\Omega_{\mathrm{d}}$ is observed initially.

When the pump power increases above $-75~\mathrm{dBm}$, the frequency comb occurs with the tooth spacing equalling $\Omega_\mathrm{m1}$, and the comb bandwidth grows steadily with higher drive power. The typical electromechanical frequency comb mediated by mechanical mode-1 is shown in the Fig.~\ref{fig:2}(b), with a fixed pump power at -29~dBm, as an example. Further increasing the pump power, higher order nonlinearity participates and results in abruptly broadened dense spectra. Analogously, when the driving is blue-detuned closed to the frequency of mechanical mode-2, i.e., $\Delta_{\mathrm{dc}}=\Omega_{\mathrm{m2}}$, the spectral evolution versus drive power is exhibited in Fig.~\ref{fig:2}(c). Under a driving power given at -25~dBm, the electromechanical frequency comb mediated by mechanical mode-2 is shown in Fig.~\ref{fig:2}(d), with the tooth spacing equalling $\Omega_\mathrm{m2}$. Extra spectral lines appear in Fig~\ref{fig:2}(c) is caused by the nonideal pump frequency, which is strictly required due to the comparatively small $g_2$ and $\gamma_2$. The more accurate $\Delta_\mathrm{dc}$ is to $\Omega_\mathrm{m2}$, the clearer the spectra would be. Similar to that in the micromechanical resonator case~\cite{yang2021asymmetric}, symmetry of frequency combs about the pump frequency (location marked as red dash line) is influenced by cavity transmission (frequency location marked as blue dash line).

\begin{figure}[tb]
\includegraphics[width=8cm]{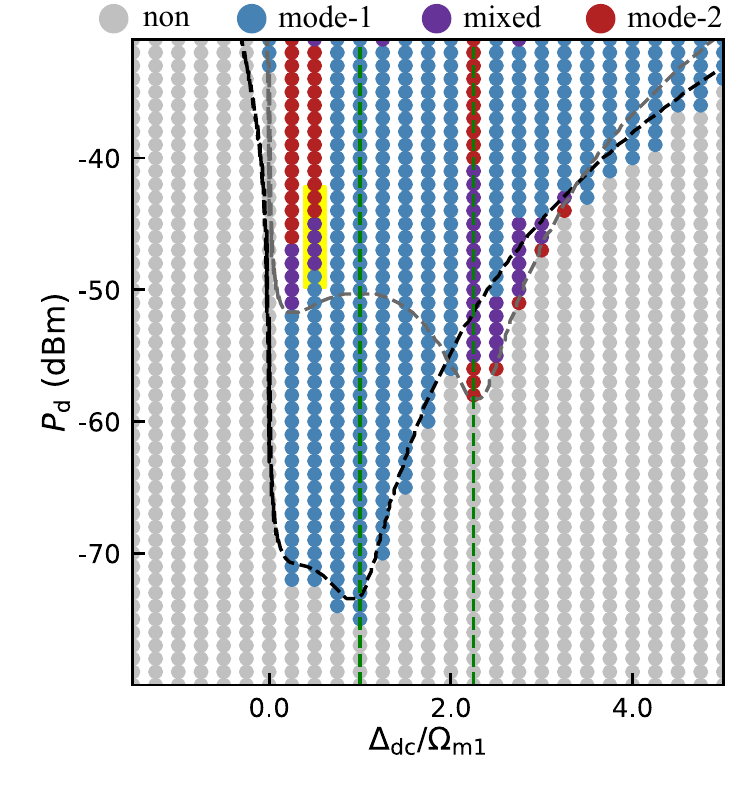}\\[5pt] 
\caption{Formation and evolution of frequency combs. Spectral responses are sorted into four types, and are marked in different colors as dots. Black (gray) dash line is single first (second) mechanical mode induced cavity field instability threshold. Two green dash lines refer to pump condition chosen for Fig.~\ref{fig:2}, and yellow highlighted region is the chosen for further analysis in Fig.~\ref{fig:4}.}
\label{fig:3}
\end{figure}

We further study dependency of the spectrum evolution on pump frequency. As shown in Fig.~\ref{fig:3}, frequency spectra with single peak at $\Omega_\mathrm{d}$ are detected initially, which are marked as gray dots. Along increased pump power, cavity field becomes unstable, and frequency combs are formed. Primary instability occurs around $\Delta_{\mathrm{dc}}=\Omega_{\mathrm{m1}}$, where the first mechanical mode influenced cavity field is most efficiently pumped. Subsequently, mode-1 mediated frequency combs are formed (blue dots in Fig.~\ref{fig:3}), and the frequency combs are similar to the spectra presented in Fig.~\ref{fig:2}(b). Similarly, mode-2 influenced cavity field instability occurs initially at $\Delta_{\mathrm{dc}}=\Omega_{\mathrm{m2}}$. Thereafter, second mechanical mode mediated frequency combs are formed (marked by red dots in Fig.~\ref{fig:3}), and the frequency combs are similar to the spectra presented in Fig.~\ref{fig:2}(d). 

The measured threshold of the pump condition (including the driving power and frequency) to form frequency combs are shown in Fig.~\ref{fig:3}. The boundary can be fitted as a combination of the black and gray dash lines. The black dash line is theoretically analysed instability threshold of the cavity field caused by the electromechanical coupling with mode-1. Meanwhile, the gray dash line marks threshold of frequency comb generation mediated by the mode-2~\cite{SM}. Overlap boundary for these two single-mechanical-mode induced instability thresholds enables transition between mode-1 mediated frequency combs and mode-2 mediated frequency combs, where hybridized spectra emerge (purple dots in Fig.~\ref{fig:3}). Teeth of the complex spectrum are unequally spaced, but can be recognized into a set of coexisting frequency combs. Such hybridization results from effective coupling between two individual mechanical modes via the cavity field~\cite{buchmann2015nondegenerate}. It is worth noting that for the most pump conditions, due to mode competition, one mechanical mode small-amplitude gains tend to suppress another~\cite{SM}. 

\begin{figure*}[tb]
\includegraphics*[width=15cm]{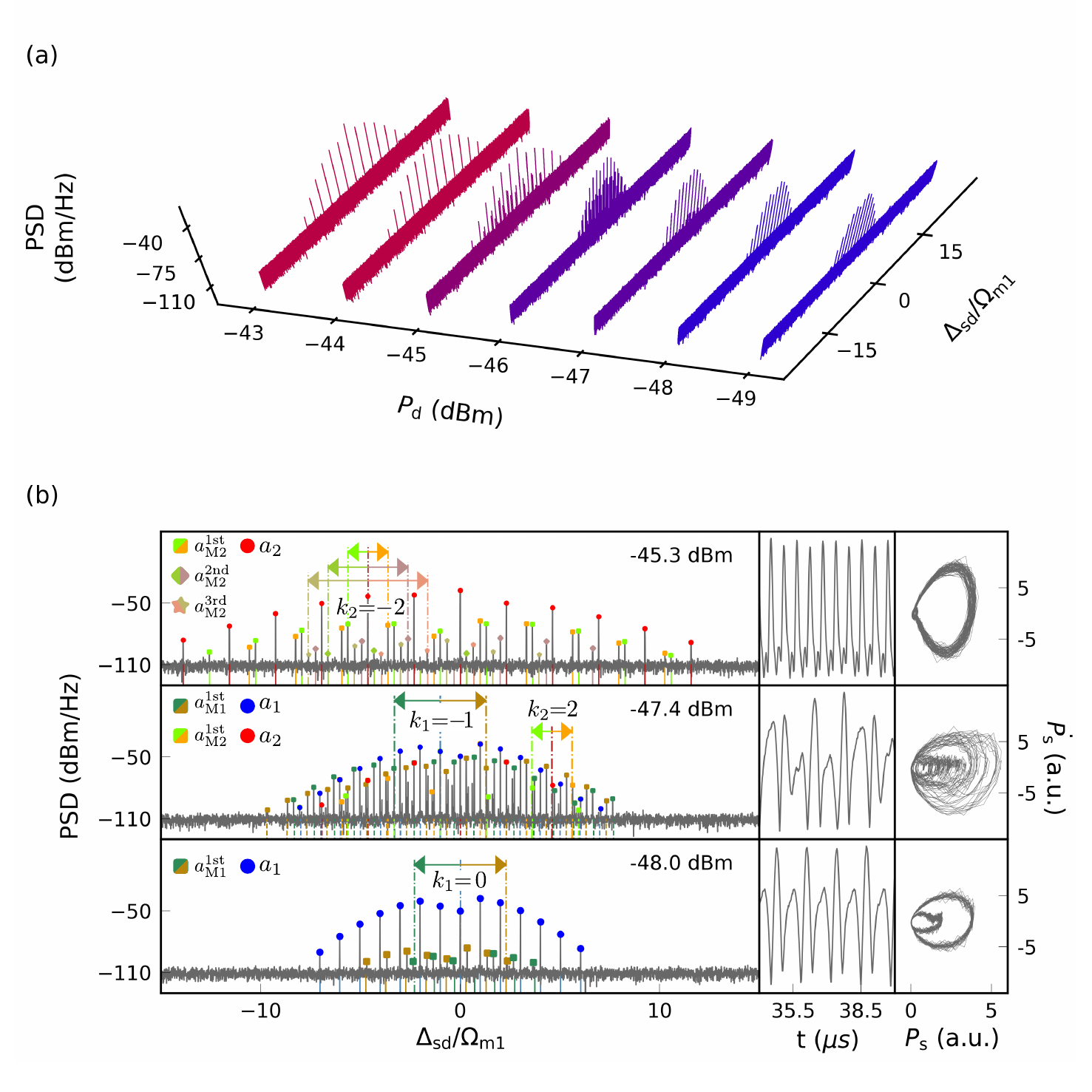}\\[5pt]
\caption{Transitional behaviours of frequency combs. (a) Pump power dependent spectrum evolution from first mechanical mode mediated frequency combs (blue ones) towards second mechanical mode mediated frequency combs (red ones), where $\Delta_\mathrm{sd}=\Omega_\mathrm{s}-\omega_\mathrm{d}$. Hybridized frequency combs (purple ones) emerge during the transition. (b) The left column exhibits the spectra within frequency domain, whose pump power are annotated. Experimental data is plotted in gray. Configurable teeth are marked in colored dots. As labelled, blue (red) round dots refer to teeth spaced in the first (second) mechanical mode frequency, i.e., $a_1(t)$ [$a_2(t)$] terms described in the text. Green and yellow square dots refer to mixed sidebands of the second order, i.e., $a_\mathrm{Mj}^\mathrm{1st}(t)$, where $\mathrm{M=s,d};~\mathrm{j=1,2}$. Similarly, diamond dots refer to $a_\mathrm{Mj}^\mathrm{2nd}(t)$, and star dots refer to $a_\mathrm{Mj}^\mathrm{3rd}(t)$. Examples of mixing procedures are picked and expressed in paired arrows, centered on single mechanical mode based tooth with order of $k_j$. The middle column of (b) exhibits the accordingly behaviour in time domain. The right column exhibits the accordingly optical trajectories in phase space, with label $P_\mathrm{s}$ representing intra-cavity field power.}
\label{fig:4}
\end{figure*}

\textit{Frequency Combs Hybridization.---}To study detailed features of frequency combs hybridization, we focus on the comb evolution when the pump detuning is fixed at $\Delta_{\mathrm{dc}}/\Omega_{\mathrm{m1}}=0.5$, and pump power varies from -49~dBm to -43~dBm (as is shown in Fig.~\ref{fig:4}, corresponding to yellow region in Fig.~\ref{fig:3}). Under this condition, evolution of frequency combs is transformed from the mode-1 determined type to the mode-2 determined type. Initially, the spectra can be expressed as $a_{1}(t)= \ensuremath{\stackrel[k_{1}=-\infty]{\infty}{\Sigma}\alpha_{k_{1}}\exp(\Omega_\mathrm{d}t)\exp(-ik_{1}\Omega_{\mathrm{m1}}t)}$. The integer label $k_{1}$ represents the order of comb teeth, with amplitude of $\alpha_{k_{1}}$, and detuning of $k_{1}\Omega_{\mathrm{m1}}$ (from the pump frequency). Note that, due to dependency of $\alpha_{k_{1}}$ on the cavity field distribution, symmetry of comb envelope about $\Omega_{\mathrm{d}}$ can be broken, especially for large $\Delta_{\mathrm{dc}}$. As the pump power increased, second mechanical mode participates and leads to sideband mixing, resulting in hybridized frequency combs, which can be expressed as

\begin{eqnarray}
a(t)=a_{1}(t)+a_{2}(t)+a_{\mathrm{s}}(t)+a_{\mathrm{d}}(t)~,
\end{eqnarray}

\noindent where $a_{2}(t)$ represents teeth detuned from the center by $k_{2}\Omega_{\mathrm{m2}}$, and can be written as $a_{2}(t)=\stackrel[k_{2}=-\infty]{\infty}{\Sigma}\alpha_{k_{2}} \exp(\Omega_\mathrm{d}t) \exp(-ik_{2}\Omega_{\mathrm{m2}}t)$. Similarly, teeth detuned by the sum and difference of several mechanical mode frequencies are $a_{\mathrm{s}}(t)=\stackrel[k_{\mathrm{s1}},k_{\mathrm{s}2}=-\infty]{\infty}{\Sigma}\alpha_{k_{\mathrm{s}}}\exp(\Omega_\mathrm{d}t)\exp[-i(k_{\mathrm{s1}}\Omega_{\mathrm{m1}}+k_{\mathrm{s2}}\Omega_{\mathrm{m2}})t]$, $a_{\mathrm{d}}(t)=\stackrel[k_{\mathrm{d1}},k_{\mathrm{d}2}=-\infty]{\infty}{\Sigma}\alpha_{k_{\mathrm{d}}}\exp(\Omega_\mathrm{d}t)\exp[i(k_{\mathrm{d1}}\Omega_{\mathrm{m1}}-k_{\mathrm{d2}}\Omega_{\mathrm{m2}})t]$. When further increase the pump power, mechanical oscillation is fully excited onto the higher mode, and thus $a(t)=a_{2}(t)$. 

Specifically, degeneracy of $k_{\mathrm{j}}$, $k_{\mathrm{sj}}$ and $k_{\mathrm{dj}}$ can be concluded as the order of mixing. For an example, when the pump condition is set to be $\Delta_\mathrm{dc}=0.5~\Omega_\mathrm{m1}$ and $P_\mathrm{d}=-48.0~\mathrm{dBm}$ [see the left lower picture of Fig.~\ref{fig:4}(b)], the spectrum can be expressed as $a(t)=a_1(t)+a_\mathrm{s1}^\mathrm{1st}(t)+a_\mathrm{d1}^\mathrm{1st}(t)$, with $a_\mathrm{M1}^\mathrm{1st}=\stackrel[k_{1}]{~}{\Sigma}\alpha_\mathrm{M_{1}}^\mathrm{1st}\exp(\Omega_\mathrm{d}t)\exp[i(k_{1}\Omega_{\mathrm{m1}}\pm \Omega_{\mathrm{m2}})t]$ for M=s,d, which represents teeth of the same spacing with $a_1(t)$ but detuned by $\pm\Omega_\mathrm{m2}$. This is a mode-1 dominated frequency comb dressed by mode-2. Phenomenally, the spectrum contains mainly teeth spaced in first mechanical mode frequency, and first order mixed sidebands centered by the $k_1$-th teeth. Subsequently, higher order mixed sidebands occur, of which the $-45.3~\mathrm{dBm}$ pumped case can be an example. This is a second mechanical mode dominated frequency comb, dressed by the first mechanical mode.

As shown in the left upper picture of Fig.~\ref{fig:4}(b), for the additional frequency comb mediated from the first order mixing procedure (marked as shortest orange and green arrows), teeth (marked as square dots) are spaced in $\Omega_\mathrm{m2}$ and detuned by $\pm\Omega_\mathrm{m1}$ from $a_2(t)$, with $k_2$ as the relative order. Further, higher order with teeth detuned by $\pm 2\Omega_\mathrm{m1}$ (marked as diamond dots) and $\pm 3\Omega_\mathrm{m1}$ (marked as star dots) call for consideration. All sum-frequency components are paired with difference terms of the same $k_2$ distribution. Total spectrum of this case can then be expressed as $a(t)=a_2(t)+\stackrel[M=s,d]{~}{\Sigma}[a_\mathrm{M2}^\mathrm{1st}(t)+a_\mathrm{M2}^\mathrm{2nd}(t)+a_\mathrm{M2}^\mathrm{3rd}(t)]$. During transformation of the two single-mechanical-mode dominated frequency comb regimes, there exist more complicated mixing procedures, e.g., with $-47.4~\mathrm{dBm}$ pumped [see the left medium picture of Fig.~\ref{fig:4}(b)], which provides finer spectrum structure. Besides, around $\Delta_{\mathrm{dc}}/\Omega_{\mathrm{m2}}=1$, similar transitional behaviour (from the mode-2 based one to the mode-1 based one) occurs. 

\textit{Conclusion.---}In this paper, we have constructed SiN-membrane based multimode superconducting cavity electromechanical device. Under cryogenic temperature, we experimentally demonstrate the evolution of frequency spectra dominated by the electromechanical nonlinearity. Threshold of cavity field instability corresponds to the formations of frequency combs. Furthermore, the frequency combs hybridization locates in two overlapped unstable boundary, i.e., where the pump detuning is near half of the frequency of mode-1, or around the frequency of mode-2. Formation for single mechanical mode mediated frequency combs and evolution towards hybridized frequency combs are both demonstrated. Specifically, teeth positions of hybridized frequency combs are analysed. At the early or late stage of the transitions, the comb teeth are spaced in frequency of one mechanical modes, and accompanied by additional mixed sidebands with detuning of integer multiple of another. 

This work could help to understand the coexistence and  hybridization of frequency combs which yields more refined frequency combs spectroscopy. Such spectroscopy exhibits mixed sidebands, accompanied by sister sidebands. As a foresight, the mode competition and hybridization provide an alternative method, besides traditional spontaneous and stimulated parametric conversion, to construct multimode entangled photon states, which provides diverse application on quantum source and quantum information processing. One prerequisite for the electromechanical microcombs entering the quantum regime is the preparation of the mechanical modes into their ground states. Introducing amplitude-modulated cooling tone, the mechanical modes with asymptotic orbits could be cooled down to their ground state when the frequency comb occurs.

This work is supported the National Key Research and Development Program of China (Grant No. 2016YFA0301200), the National Natural Science Foundation of China (Grant No.62074091, No.12004044, and U1930402), China Postdoctoral Science Foundation (Grant No. 2021M700442), and the Science Challenge Project (Grant No.TZ2018003). 

S.~W. and Y.~L. contributed equally to this work.

\end{document}